\documentclass[twocolumn, showpacs,preprintnumbers,amsmath,amssymb]{revtex4}
\usepackage{amssymb}
\usepackage{xcolor}
\usepackage{graphicx}% Include figure files
\usepackage{dcolumn}% Align table columns on decimal point
\usepackage{bm}% bold math
\usepackage{multirow}
\usepackage{booktabs}
\usepackage{longtable}
\usepackage{natbib}
\usepackage{lscape}
\usepackage{soul}
\usepackage{amssymb}
\usepackage{amsmath}	% Advanced maths commands
\usepackage{color}
\usepackage{cases} 

\usepackage{array}

\begin{document}
\preprint{APS/123-QED}

\title{Dissociative recombination and vibrational excitation of BF$^{+}$ in low energy electron collisions}
\author{J. Zs Mezei$^{1,2,3,4}$}\email[]{mezei.zsolt@atomki.hu}
\author{F. Colboc$^{1}$}
\author{N. Pop$^{5}$}
\author{S. Ilie$^{1,5}$}
\author{K. Chakrabarti$^{1,6}$}
\author{S. Niyonzima$^{7}$}
\author{M. Leppers$^{2}$}
\author{A. Bultel$^{8}$}
\author{O. Dulieu$^{2}$}
\author{O. Motapon$^{9}$}
\author{J. Tennyson$^{10}$}
\author{K. Hassouni$^{4}$}
\author{I. F. Schneider$^{1,2}$}%\email[]{ioan.schneider@univ-lehavre.fr}
\affiliation{$^{1}$LOMC CNRS$-$Universit{\'{e}} du Havre$-$Normandie Universit{\'{e}}, 76058 Le Havre, France}
\affiliation{$^{2}$LAC, CNRS$-$Universit\'e Paris-Sud$-$ENS Cachan$-$Universit\'e Paris-Saclay, 91405 Orsay, France}%
\affiliation{$^{3}$HUN-REN Institute for Nuclear Research (ATOMKI), H-4001 Debrecen, Hungary}%
\affiliation{$^{4}$LSPM, CNRS$-$Universit\'e Paris 13$-$USPC, 93430 Villetaneuse, France}%
\affiliation{$^{5}$Fundamental of Physics for Engineers Department, Politehnica University Timisoara,  300223 Timisoara, Romania}%
\affiliation{$^{6}$Dept. of Mathematics, Scottish Church College, Calcutta 700 006, India}
\affiliation{$^{7}$D\'ept. de Physique, Facult\'e des Sciences, Universit\'e du Burundi, B.P. 2700 Bujumbura, Burundi}
\affiliation{$^{8}$CORIA CNRS$-$Universit\'{e} de Rouen$-$Universit{\'{e}} Normandie, F-76801 Saint-Etienne du Rouvray, France}%
\affiliation{$^{9}$LPF, UFD Math., Info. Appliq. Phys. Fondamentale, University of Douala, P. O. Box 24157, Douala, Cameroon}%
\affiliation{$^{10}$Dept. of Physics and Astronomy, University College London, WC1E 6BT London, UK}%
\date{\today}

\begin{abstract}
 The latest molecular data - potential energy curves and Rydberg-valence interactions - characterising the super-excited electronic states of BF are reviewed in order to provide the input for the study of their fragmentation dynamics. Starting from this input, the main paths and mechanisms of BF$^+$ dissociative recombination and vibrational excitation are analysed. Their cross sections are computed for the first time using a method based on the multichannel quantum defect theory (MQDT), and Maxwellian rate-coefficients are calculated and displayed in ready-to-be-used format for low temperature plasma kinetics simulations.
\end{abstract}

% PACS, the Physics and Astronomy
                             % Classification Scheme.
%\keywords{coupled-channel, optical shielding, KCs}%Use showkeys class option if keyword

\maketitle

\section{Introduction}\label{sec:intro}

Boron fluoride (BF$_3$) containing plasmas are important for
applications in the field of material processing. BF$_3$ and
Ar$/$BF$_3$ plasmas are for instance used for p-type doping in the
semi-conductor industry~\cite{torregrosa2004,matsui2004}. Moreover,
electrical discharges generated in complex mixture of BF$_3$ with
nitrogen containing compounds have been proposed and investigated for
the deposition of boron nitride coatings~\cite{yu2003,yamamoto2006}.
All these processes make use of low pressure high density plasma
sources where BF$_3$ interacts with the free electrons of the plasma
through elastic and inelastic collisions. These collisions are not
only the driver for the generation of the active key-species of the
process, e.g., the positive ions for the plasma immersion ion
implantation (PIII) doping processes or active radicals for the
deposition processes, but they also contribute and govern the
discharge equilibrium and stability through ionisation, attachment and
recombination processes ~\cite{agarwal2007,farber1984,kim2000}. In
particular, for high density magnetised inductively coupled
radio-frequency sources used in many of these processes, the surface
losses are limited due to the magnetic confinement, which makes
the contribution of ion recombination and mutual recombination
processes to the discharge equilibrium very significant. In this
context, the BF$^+$ ion can represent a significant fraction of the ions
produced when working at high power density discharge conditions for
which BF$_3$ dissociation and dissociative ionisation are strongly
enhanced and the plasma is dominated by BF$_2$ and BF
fragments~\cite{maury2016}. These species can therefore significantly
contribute either as an active species in the processes or in the
discharge ionisation-recombination equilibrium. The investigation of
the dissociative recombination of BF$^+$ is therefore of importance
for understanding BF$_3$ plasma processes.

The kinetic description of all the above mentioned environments is
based on the knowledge of the rate coefficients of the dominant
reactions, including those between electrons and molecular ions.  As
for the BF$^{+}$ ions, their abundance and their vibrational
distribution are strongly affected by the dissociative recombination
(DR),

\begin{equation}
 \mbox{BF}^{+}(v_{i}^{+}) + e^{-}\rightarrow  \mbox{B} + \mbox{F},
\label{eq:dr}
\end{equation}

\noindent
and also by other competitive processes, such as inelastic
collisions ($v_{f}^{+}>v_{i}^{+}$) and super-elastic collisions
($v_{f}^{+}<v_{i}^{+}$) 
which are also known as collisions
of the second kind:

\begin{equation}
 \mbox{BF}^{+}(v_{i}^{+}) + e^{-}\rightarrow  \mbox{BF}^{+}(v_{f}^{+}) + e^{-},
\label{eq:diff}
\end{equation}
\noindent
where $v_{i}^{+}$ and $v_{f}^{+}$ stand for the initial and final
vibrational quantum number of the target ion, and rotational structure
is neglected.

In the present work, we compute the DR and vibrational transition (VT)
- vibrational excitation/de-excitation (VE/VdE) - cross sections and
rate coefficients for the lowest three vibrational levels of the
BF$^+$ in its ground electronic state using the multichannel
quantum defect theory (MQDT).
The paper is structured as follows: Section~\ref{sec:theory} outlines the main ideas and steps of our MQDT approach. Section~\ref{sec:moldata} presents the molecular data used in the calculation. The main results are given in section~\ref{sec:res} and the paper ends by conclusions. 

\section{The MQDT-type approach to DR}{\label{sec:theory}}

The MQDT approach~\cite{seaton83,gj85,jungen96,giusti80} has been shown to be a powerful method for the evaluation of the cross sections in collisions of electrons/photons with molecular cation/neutral
systems. It was applied with great success in calculating DR cross sections to several diatomic systems like H$_2^+$ and its isotopologues \cite{giusti83,ifs-a09,takagi1993,tanabe95,ifs-a18,amitay99}, O$_2^+$ \cite{ggs91,guberman-dr99}, NO$^+$~\cite{sn90,ifs-a22,jt260},
LiH$^{+}$ \cite{cGreene2007}, HeH$^{+}$ \cite{haxtonGreene2009}, LiHe$^{+}$ \cite{curikGianturco2013} 
and triatomics like H$_{3}^{+}$ \cite{ifsa26,kokoou01,kokoou03}, and for its competitive processes like ro-vibrational transitions in case of NO$^+$~\cite{motapon06b}, CO$^+$~\cite{mezei2015} and H$_2^+$~\cite{Ngassam1,epee2015}. Recently, a global version of MQDT~\cite{jungen2011} has been used to describe  the photoabsorption, photoionisation and photodissociation of H$_2$~\cite{mezei2012,mezei2014,sommavilla2016}, providing very good agreement with 
highly accurate experimental results.

In the present paper, we use an MQDT-type method to study the electron-impact collision processes given by eqs.~(\ref{eq:dr}) and (\ref{eq:diff}) which result from the quantum interference of the {\it direct} mechanism - the capture takes place into a dissociative
state of the neutral system (BF$^{**}$) - and the {\it indirect} one -  the capture occurs \textit{via} a Rydberg state of the molecule BF$^{*}$ which is  predissociated by the BF$^{**}$ state. In both mechanisms the autoionization is in competition with the predissociation and leads, through the  reaction (\ref{eq:diff}), to {\it super-elastic collision} (SEC) ($v^+_i>v^+_f$ in eq.~(\ref{eq:diff})), {\it elastic collision} (EC) ($v^+_i=v^+_f$) and {\it inelastic collision} (IC) ($v^+_i<v^+_f$). 

A detailed description of our theoretical approach is given in previous studies \cite{jt591,mezei2015}.
The major steps of the method only can be briefly outlined as follows:
\begin{enumerate}
\item {\it Defining the interaction matrix} $\boldsymbol{\mathcal{V}}$:\\
Within a quasi-diabatic representation of the BF states, the interaction matrix is based on the computed \cite{chakrabarti09,chakrabarti11} couplings between 
\textit{ionisation} channels - 
	associated to
	the vibrational levels $v^+$ of the cation  
	and 
	to the orbital quantum number $l$ of the incident/Rydberg electron - 
and 
\textit{dissociation} channels $d_j$.

\item {\it Computation of the reaction matrix} $\boldsymbol{\mathcal{K}}$:\\ 
Given $\boldsymbol{H_0}$ the Hamiltonian of the molecular system under study in which the Rydberg-valence interaction is neglected, we adopt  the second-order perturbative solution
for the Lippman-Schwinger integral equation \cite{Ngassam1,Florescu2003,motapon06b}, 
written in operatorial form as:
\begin{equation}\label{eq:solveK}
\boldsymbol{\mathcal{K}}= \boldsymbol{\mathcal{V}} + \boldsymbol{\mathcal{V}}{\frac{1}{E-\boldsymbol{H_0}}}\boldsymbol{\mathcal{V}}.
\end{equation}
\noindent

\item {\it Diagonalization of the reaction matrix,} \\ 
yields the corresponding eigenvectors and eigenvalues which are used to build the  eigenchannel wavefunctions.

\item {\it Frame transformation from the Born-Oppenheimer (short-range) to the close-coupling (long-range) representation,}\\ 
relying, for a given electronic total angular momentum quantum number $\Lambda$ 
and a given orbital quantum number of the incident/Rydberg electron $l$, on the
quantum defect $\mu_{l}^{\Lambda}(R)$ and on the eigenvectors and eigenvalues of the K-matrix.

\item {\it Construction of the generalised scattering 
matrix $\boldsymbol{{X}}$}, \\ 
based on the frame-transformation coefficients, 
this matrix  being organised in blocks associated to open and/or closed ($o$ and/or $c$ respectively) channels:

\begin{equation}
\qquad
\boldsymbol{{X}}=
 \left(\begin{array}{cc} \boldsymbol{X_{oo}} & \boldsymbol{X_{oc}}\\
                   \boldsymbol{X_{co}} & \boldsymbol{X_{cc}} \end{array} \right).
\end{equation}

\item {\it Construction of the generalised scattering matrix $\boldsymbol{\mathcal{S}}$}, \\ 
\begin{equation}\label{eq:solve3}
\boldsymbol{S}=\boldsymbol{X_{oo}}-\boldsymbol{X_{oc}}\frac{1}{\boldsymbol{X_{cc}}-\exp(-i2\pi\boldsymbol{ \nu})}\boldsymbol{X_{co}}.
\end{equation}
\noindent
based on the open channels, the first term in eq.~({\ref{eq:solve3}}), but also on their mixing with the closed ones, given by the second term, the denominator being responsible for the resonant patterns in the shape of the cross section
\cite{seaton83}. Here the matrix $\exp(-i2\pi\boldsymbol{ \nu})$ is diagonal and contains the effective quantum numbers $\nu_{v^{+}}$ associated to the vibrational thresholds of the closed ionisation channels.

\item {\it Computation of the cross-sections:}\\
For each of the relevant BF states, which are grouped by symmetry properties: 
electronic total angular momentum quantum number $\Lambda$, electronic spin singlet/triplet,
and for a given target cation vibrational level $v_i^+$ and energy of the incident electron $\varepsilon$, the dissociative recombination and the vibrational excitation/de-excitation cross sections are computed using, respectively:

\begin{equation}\label{eqDR}
\sigma _{{\rm diss} \leftarrow v_{i}^{+}}^{{\rm sym}}=\frac{\pi}{4\varepsilon} \rho^{{\rm sym}}\sum_{l,j}\mid S_{d_{j},lv_{i}^{+}}\mid^2,
\end{equation}
\begin{equation}\label{eqVe}
\sigma _{v_{f}^{+} \leftarrow v_{i}^{+}}^{{\rm sym}
}=\frac{\pi}{4\varepsilon} \rho^{{\rm sym}}\sum_{l,l'}\mid S_{l' v_{f}^{+},lv_{i}^{+}}-\delta_{l'l}\delta_{v_{i}^{+}v_{f}^{+}}\mid^2,
\end{equation}
\noindent
where $\rho^{{\rm sym}}$ stands for the ratio between the multiplicity of the involved electronic states of BF and that of the target, BF$^+$.
\end{enumerate}

\section{Molecular data}{\label{sec:moldata}}
The molecular data necessary to model the DR are the
potential energy curve (PEC) of the ground state of the ion, the PECs of the 
neutral valence dissociative states interacting with the ionization continua, those of 
the Rydberg states associated to these continua below the threshold, and 
all the relevant Rydberg-valence couplings.
These molecular data
are mostly obtained from {\it ab initio} R-Matrix calculations. In order to extend the PECs to
small and large values of the internuclear distance R, we  used
the quantum chemistry molecular data of \cite{magoulas13}. 

\subsection{\textit{Ab initio R-Matrix calculations for the molecular states}}
\label{subsec:disspec}

{\it Ab initio} R-matrix calculations of electron collision with the
BF$^+$ ion yielding the relevant molecular data were performed by
three of us \cite{chakrabarti09,chakrabarti11}. The R-matrix method is
a state-of-the-art quantum scattering technique which can be used to
calculate the properties of bound and resonant electronic states of a
molecule \cite{tennyson10}. The BF$^+$ target states were first
obtained by performing a configuration interaction (CI) calculation.
Subsequently this CI target wave function was used in an R-matrix
calculation.  The bound states of the BF molecule were obtained by
searching for negative energy solutions using the BOUND program
available within the R-Matrix code suite \cite{jt106,carr12}.
This also produces quantum defects of the bound states.
Resonances were detected and fitted to a Breit-Wigner profile
\cite{jt31} to obtain their energies and widths. Subsequently the
electronic couplings $V_{d_j,l}$ were obtained from the resonance
widths $\Gamma_{d_j,l}$ using the relation
\begin{equation}{\label{width}}
V_{d_j,l} = \sqrt{\frac{\Gamma_{d_j,l}}{2\pi}}. 
\end{equation}
The calculations were repeated for $11$ internuclear distances in the
range $1.5~a_0 - 3.5~a_0$. 
While the R-matrix calculations used
several partial waves to represent the scattering, one single  
{\em global} (i.e. not {\em l}-resolved) resonance width was 
produced for each symmetry. 
Consequently, one generic partial wave for each symmetry, corresponding to the highest effective quantum number produced by the R-matrix computation
(see Figure 3 from~\cite{chakrabarti11}),
was considered to take in charge the total strength of the Rydberg-valence interaction
in the current calculations. 
Dissociative curves were constructed using the resonance
data above the ion PEC and their continuation as bound states below
the ion PEC.

\begin{figure*}[t]
\centering
\includegraphics[width=0.7\textwidth]{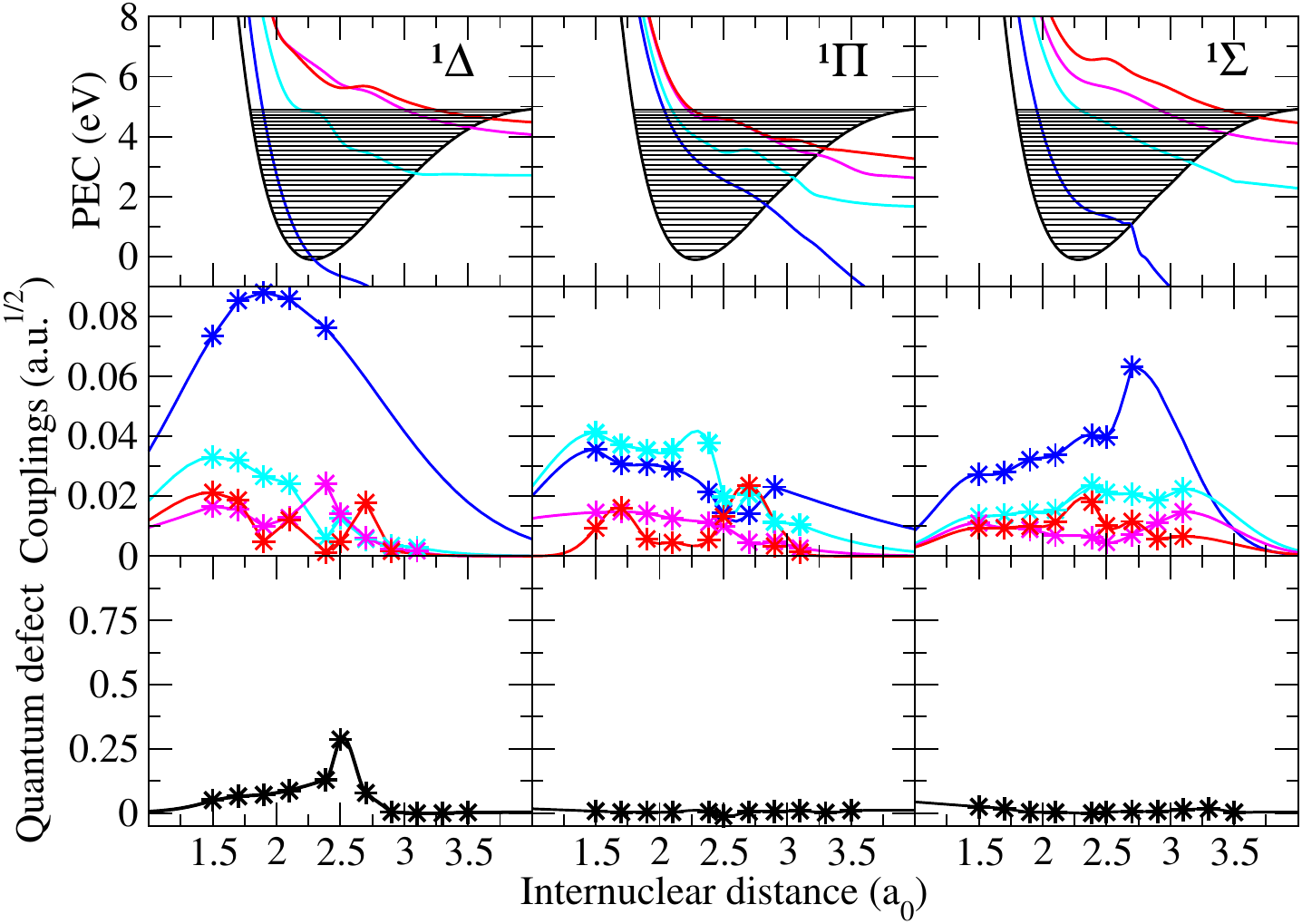}
\caption{
The dissociative states, electronic couplings and quantum defects relevant for BF$^+$ dissociative recombination within the major symmetries, indicated in each figure. Top panel: the potential energy curves of the dissociative BF$^{**}$ states (colour, cf section~\ref{sec:moldata}), the ground state potential energy curve of  BF$^{+}$ and its vibrational levels (black). Mid panel: couplings between the valence BF$^{**}$ dissociative states and the ionisation continua BF$^{+}$ + e$^-$. Bottom panel: quantum defects 
characterising
the Rydberg series for the given symmetry. The symbols stand for the R-matrix data points.
}
\label{pecs}
\end{figure*}

\subsection{\textit{Modeling of the potential energy curves and electronic 
couplings}}\label{subsec:disspec}

Our earlier studies on molecular systems such as H$_2$~\cite{epee2015}, HD~\cite{motapon14}, NO~\cite{motapon06b}, N$_2$~\cite{jt591} and CO~\cite{mezei2015} revealed that accurate and complementary R-matrix
and quantum chemistry calculations are needed for a reliable description of the molecular dynamics in electron/molecular cation collisions. In particular, we found that the DR cross section is extremely sensitive to the position of the PECs of the neutral dissociative states with respect to that of the target ion. Indeed,  a slight  change of the crossing point of the PEC of a neutral dissociative state with that of the ion ground state can lead to a significant change in the predicted DR cross section. In addition, the PECs of the dissociative states must also go to the correct asymptotic limits for large values of the internuclear distance $R$. 

In order to fulfil these accuracy criteria, we smoothly matched the
PECs corresponding to R-matrix resonances, situated above the ion PEC,
to branches of adiabatic PECs coming from previous quantum chemistry
calculations \cite{magoulas13}. Some of these latter PECs allowed us
to complete the relevant diabatic landscape. More specifically, since
for the states having $^1\Sigma^+$ and $^1\Pi$ symmetry the avoided
crossings involved always two adiabatic states only, the adiabatic
PECs can be a transformed into (quasi-)diabatic PECs using a
$2\times2$ rotation matrix~\cite{roos2009}. Meanwhile, far from the
avoided crossing, we assumed that the adiabatic and diabatic potential
curves are identical. This assumption forces the rotational angle to
be a function that varies steeply from $0$ to $\pi/2$, chosen in the
present case a tangent hyperbolic function. In this way we were able
to extract the diabatic PECs for the dissociative states with singlet
symmetries relevant for DR.

The electronic couplings provided by the R-matrix theory
\cite{chakrabarti11} for the triplet states are at least one or even
two orders of magnitude smaller then those of the singlet ones, so
they are omitted in the present calculation.

The data on the PECs are limitted to 
internuclear distances with $R \leq 8$~a$_0$. We completed the diabatic PECs by
adding a long-range tail of $D-\sum_{n}C_{n}/R^{n}$ type
\cite{stone1996,maxence2011} for larger values of $R$, where $D$ is
the dissociation limit as given in the \cite{nistdatabase}. For the
ground state of BF$^{+}$, the leading term $C_{3}/R^{3}$ of the
multipolar expansion comes from the interaction between the charge of
B$^{+}$ and the quadrupole moment of fluorine in its ground state
$^{2}P^{o}(M_{L}=0)$, with $M_{L}$ the azimuthal quantum number of $F$
with respect to the internuclear axis. The corresponding coefficient
$C_{3}=Q=0.731$~a.u. ~\cite{medved2000} gives rise to a repulsive
interaction, which is balanced by the next term $C_{4}/R^{4}$ of the
multipolar expansion, due to the polarisation of the electronic cloud
of $F$ by the ion $B^{+}$. The coefficient is given by
$C_{4}=-\alpha_{zz}/2=-1.72$~a.u., where $\alpha_{zz}=3.43$~a.u.
\cite{medved2000} is the static dipole polarisability of the ground
state $^{2}P^{o}(M_{L}=0)$ of fluorine. As for the states of the BF ,
the $C_{n}$ coefficients are calculated as fitting parameters to
ensure a smooth connection with the existing part of the PECs.

The top panel of Figure \ref{pecs} shows the PEC of the ion ground state - together with its vibrational structure - and the PECs of the DR-relevant dissociative states obtained using the procedure outlined above and taken into account in the present MQDT calculation. The electronic couplings of these latter states with the ionization continua, computed from the R-Matrix-produced autoionization widths followed by Gaussian-type extrapolations to small and large internuclear distances, are shown in the mid panel of the same figure. The bottom panel gives the quantum defects characterising the Rydberg series for the different 
symmetries. They are small in absolute value and slowly variable with the internuclear distance $R$, with a notable exception in
a small region around 2.5~a$_0$ in the case of the $^1\Delta$ symmetry, due to local 
relatively strong Rydberg-valence interactions.

\section{Results and discussion}\label{sec:res}
\subsection{Evaluation of the cross section using the MQDT approach}

Using the set of molecular data (PECs, electronic couplings and
quantum defects) determined as described in the previous section, we
performed a series of MQDT calculations of cross sections for DR
and competitive processes, assuming BF$^+$ to be initially in its
electronic ground state X\ $^2\Sigma^+$ and on one of its lowest
vibrational levels $v_i^+=0, 1$ and $2$. We have considered that the
reactive processes take place via BF-states of total symmetry of
$^1\Delta$, $^1\Pi$, $^1\Sigma^+$ - see Figure \ref{pecs} - and we
neglected rotational and spin-orbit effects.

We consider incident electron energies from $0.01$ meV up to $5$ eV
- the dissociation energy of BF$^+$ X\ $^2\Sigma^+$ electronic state -
and all the $30$ vibrational levels of the ion are included in
the calculations.

At very low energy, all the excited levels are associated with closed
ionisation channels, as defined in section~\ref{sec:theory},
responsible for temporary resonant capture into Rydberg states. As the
energy increases, more and more ionisation channels open, which
results in autoionization, leading to competitive processes, such as IC
and SEC, and decreasing the flux of DR.

\begin{figure}[t]
\centering
\includegraphics[width=0.95\columnwidth]{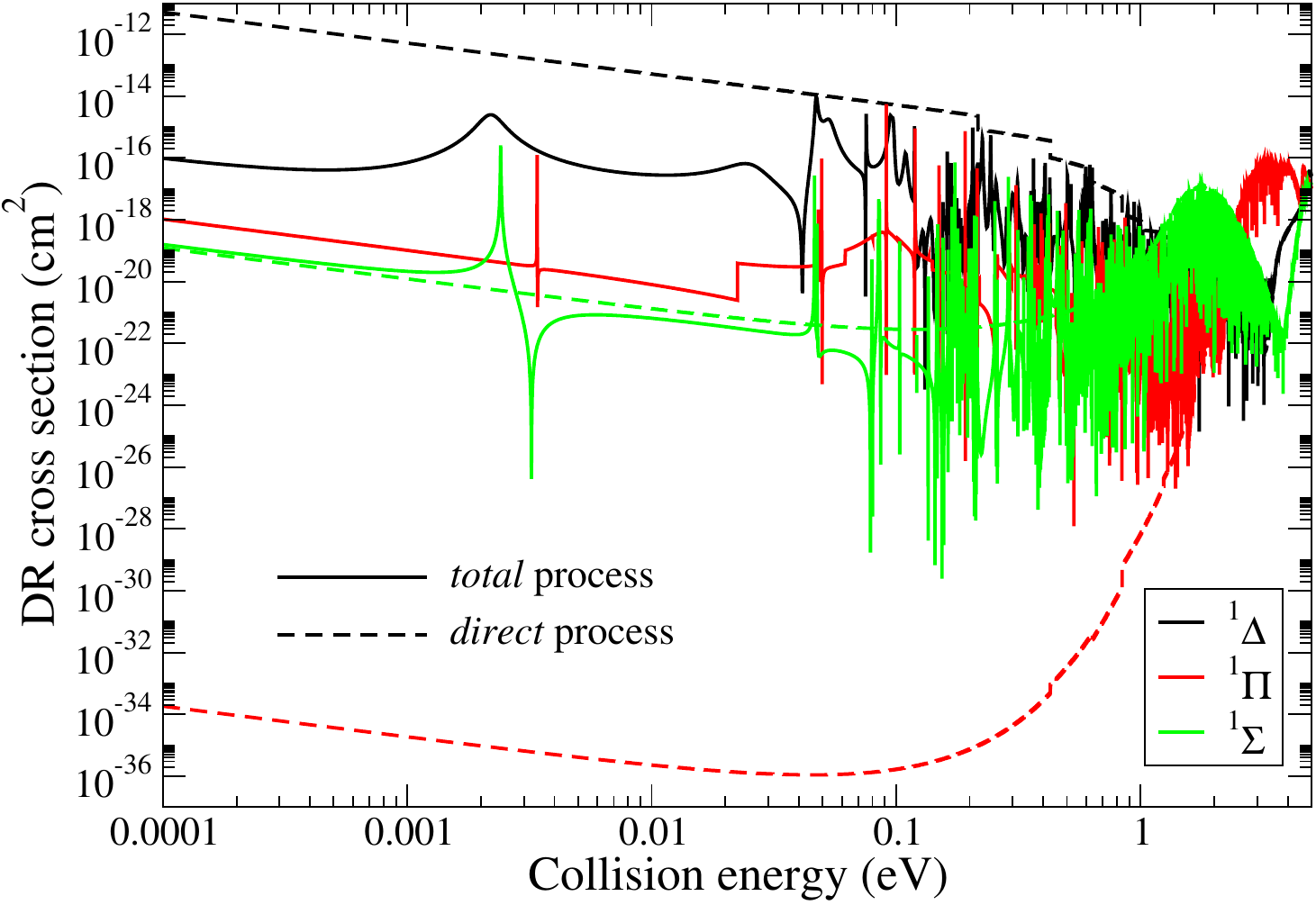}
\caption{DR cross section for the direct (dashed lines) and the total (direct and indirect) processes (continuous lines) for each symmetry, with the ion initially in $v_i^+=0$, 
}
\label{xs_sym}
\end{figure}

The direct electronic couplings between ionisation and dissociation
channels - mentioned in paragraph (i) of section~\ref{sec:theory} -
were extracted from the autoionization widths (see
eq.~(\ref{width})) of the valence states calculated by
\cite{chakrabarti09}.  For each
dissociation channel available, we  considered interaction with
the most relevant series of Rydberg states including only one partial
wave for each symmetry.

\begin{figure}[t]
\centering
\includegraphics[width=0.95\columnwidth]{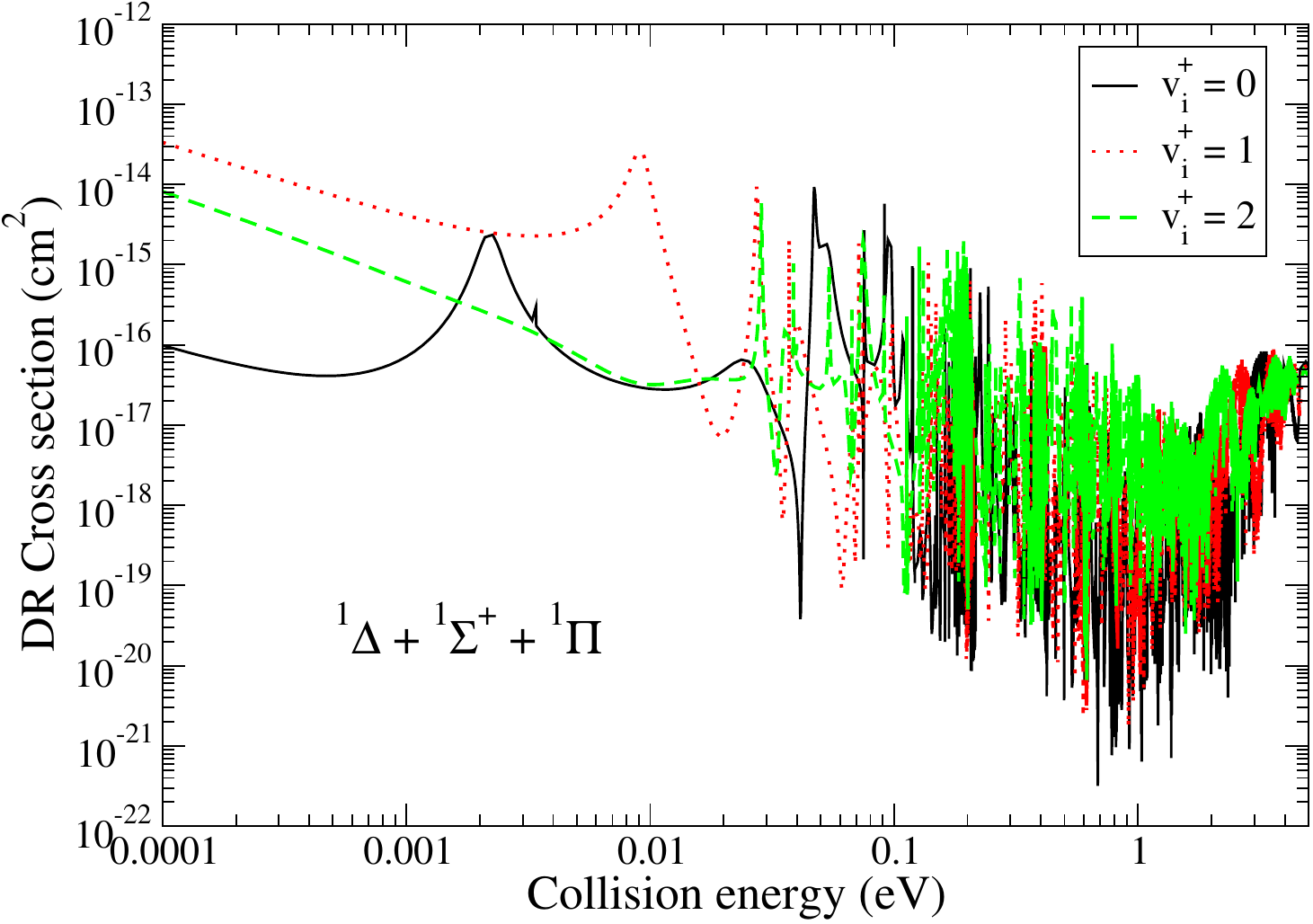}
\caption{
Total DR cross sections for different initial vibrational levels of the ion, for all of the symmetries considered. Black continuous lines are for DR from $v_i^+=0$, while the red dotted and green dashed lines are for DR from  $v_i^+=1$ and $2$, respectively.
}
\label{xs_sum}
\end{figure}

By examining the magnitude of the valence-Rydberg couplings (shown in
the bottom panel of figure~\ref{pecs}) and the positions of the
crossing points between the dissociative states correlating to the
B$(2p^1)$ $+$ F$(2p^5)$ atomic limits with the ionic ground state -
blue and black curves in the first figure of the upper panel of the
same figure - at low collision energies (up to $1$ eV) one may predict
that the major part of the total cross section comes from the
$^1\Delta$ symmetry (solid black line in figure~\ref{xs_sym}), while
the $^1\Sigma$ and $^1\Pi$ symmetries (solid green and red lines in
the same figure) only make a minor contribution to the cross section.
The importance of these latter two symmetries is revealed at higher
collision energies, when the crossing of the first dissociative state
with the molecular ion becomes favorable. At collision energies about
$1$ eV, the $^1\Sigma^+$ states give a larger contribution to the DR
cross section than the $^1\Delta$ symmetry, which in turn is exceeded
by the $^1\Pi$ symmetry at about $2$ eV. Moreover, one can notice that
at higher collision energies, around $3.5$ eV, when all the
dissociative states are open, the DR cross section shows a sharp
revival and, at $5$ eV, the resonance structures disappear since all
the ionization channels are open, and consequently the direct
process only drives the DR.

Figure~\ref{xs_sym} shows the importance of the {\it indirect} process for the DR of vibrationally relaxed BF$^+$ at low energies. 

Within the $^1\Delta$ symmetry, the entrance ionization channel associated with the  $v_i^+=0$ state is strongly coupled to the dissociation continuum (see Figure 1),  allowing the direct process to dominate. The indirect process diminishes the cross section - {\it destructive} interferences with the direct one - due to stronger coupling between closed channels associated with highly-excited vibrational levels - $v^+=12-16$ - and the dissociative ones.

Within the $^1\Pi$ symmetry one finds the opposite. The direct process
gives only a minor contribution to the DR cross section, due to the
poor Franck-Condon overlap between the vibrational wave function of
the initial state of the ion and the dissociative neutral states.
However, the much stronger couplings between some closed ionisation
channels - $v^+=8-10$ - and the dissociative states significantly
increases the total cross section through the indirect
mechanism~\cite{pop2012,schneider2012} resulting into {\it constructive} quantum interferences.

Finally, for the states with $^1\Sigma^+$ symmetry, the indirect
process plays a relatively minor role, the magnitude of the total
cross section being given mainly by the direct process.  The indirect
process is responsible for resonant structures which are a
consequences of both constructive and destructive interference.

\begin{figure}[t]
\centering
\includegraphics[width=0.95\columnwidth]{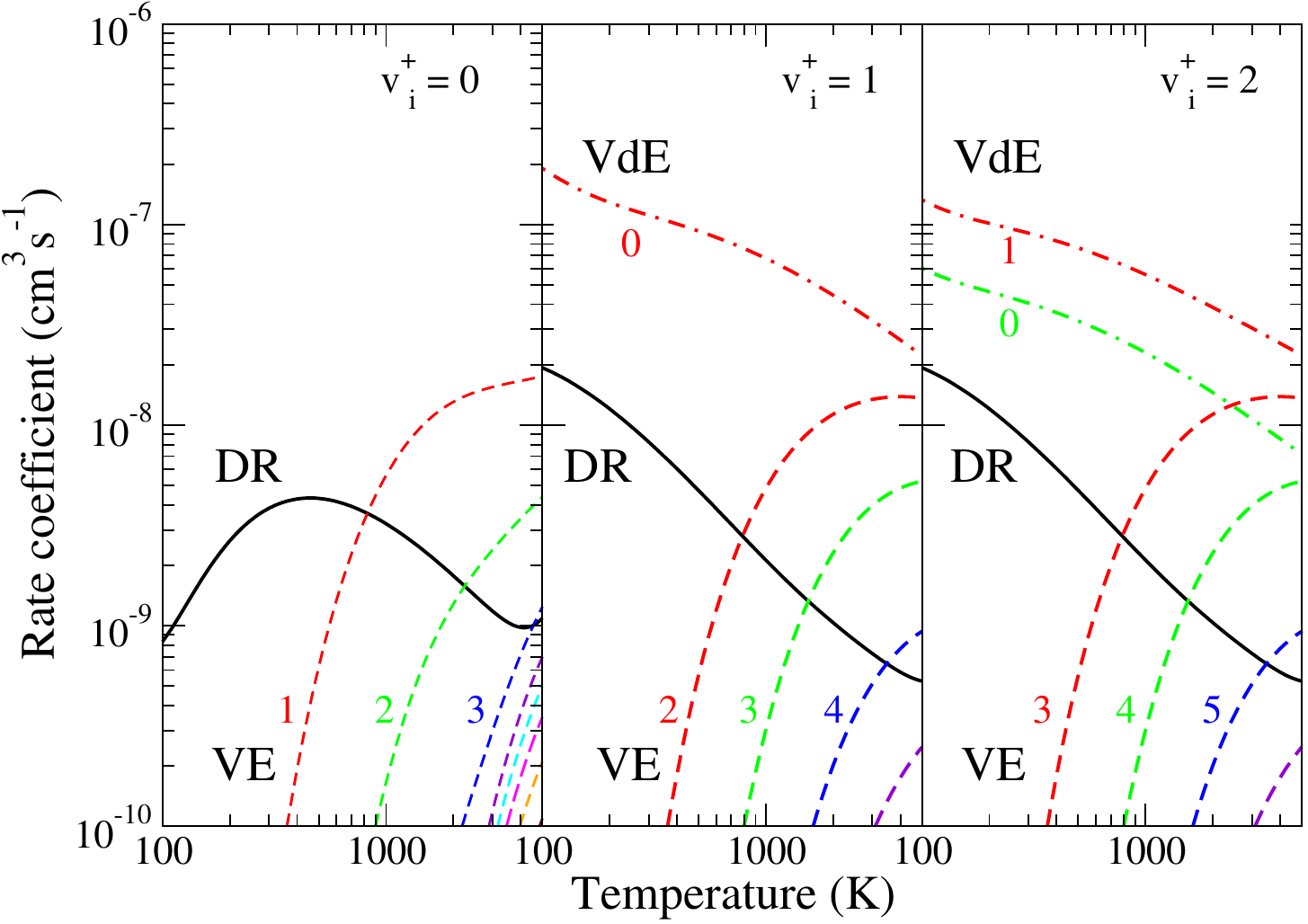}
\caption{
DR and state-to-state VE and VdE 
Maxwell rate coefficients of BF$^+$ in its ground electronic state,  $v_i^+$ standing for the initial vibrational quantum number of the target ion. Curves of the same color show the rate coefficients for the vibrational (de-)excitations corresponding to the same $|\Delta v|=|v^+_f-v^+_i|$; $v^+_f > v^+_i$ for the VE and $v^+_f <v^+_i$ for the VdE global rate coefficients. The final vibrational quantum numbers of the ion are indicated for these processes.}
\label{isorates_all}
\end{figure}

For a given initial vibrational level of BF$^+$, the total DR cross
section is obtained by summing over the partial cross sections of
the three symmetries contributing to the process - $^1\Delta$, $^1\Pi$
and $^1\Sigma^+$ - given in eq,~(\ref{eqDR}).  The results of the
calculations performed for the lowest three vibrational levels of the
ground electronic state of the ion are shown in Figure \ref{xs_sum}.

\subsection{Reaction rate coefficients for DR and its competitive processes}

Besides the dissociative recombination cross sections presented in
Figure~\ref{xs_sum}, we use eq.~(\ref{eqVe}) to 
calculate the vibrational excitation (VE) and de-excitation
(VdE) cross sections for the same vibrational levels of the target.
Since many of the features of the VE and VdE cross sections either are
similar to, or may be understood with the same reasoning as those of DR,
we do not display them here.  Instead, we present in this section
their Maxwell rate coefficients for a broad range of electronic
temperatures, in comparison with those of the DR. All these data are
relevant for cold non-equilibrium plasmas.

Figure \ref{isorates_all} shows the total DR and vibrational
transition - VE and VdE - rate coefficients for the three lowest
vibrational levels of the ion.  The DR rate for $v_i^+=0$ shows
different behaviour to those for the $v_i^+=1$ and $2$ states.  It has
a maximum at an electron temperature $T_e$ of about 400 K, while the rates
for the other two states are monotonically decreasing with  $T_e$,
reaching nearly the same magnitude at higher temperatures.
This behaviour for $v_i^+=0$ comes from the strong resonant peaks just
below 0.01 eV, see Figure 2 and black curve in Figure 3.

For other recently studied molecular systems, N$_2$~\cite{jt591} and
CO~\cite{mezei2015}, the DR clearly dominates the electron-ion
collisions. However, for BF the calculated VdE rate coefficients are
larger than the DR-ones, and VE rate coefficients exceed the DR-ones
at high $T_e$ by almost an order of magnitude.  Thus, one can conclude
that in the case of the BF$^+$ containing plasma, the internal
vibrational energy is much more important for  energy-exchange via VE
and VdE, than the kinetic energy release via the exothermic DR.

To facilitate the use of our rate coefficients (Figure
\ref{isorates_all}) in modeling calculations, we fit them 
using modified Arrhenius-type formulas. The calculated DR rate
coefficients for $v=0, 1, 2$ are given by :

\begin{equation}
\label{eqfit_DR}
k^{DR}_{BF^+,v}(T_e)=A_{v}{T_e}^{\alpha_{v}}\exp{\left[-\sum_{i=1}^7\frac{B_{v}(i)}{i\cdot{T_e}^i}\right]}
\end{equation}

\noindent over the electron temperature range $100$ K $<T_e<3000$ K. The parameters $A_v$, $\alpha_{v}$, and B$_v$(i) are listed in table \ref{fit_DR}. 
The corresponding formula for the vibrational transitions (VE and VdE) has the form:

\begin{equation}
\label{eqfit_VT}
k^{DR}_{BF^+,v'\rightarrow v''}(T_e)=A_{v'\rightarrow v''}{T_e}^{\alpha_{v'\rightarrow v''}}\exp{\left[-\sum_{i=1}^7\frac{B_{v'\rightarrow v''}(i)}{i\cdot{T_e}^i}\right]}
\end{equation}

\noindent over the electron temperature range $T_{min}<T_e< 3000$ K. The parameters T$_{min}$, $A_{v'\rightarrow v''}$, $\alpha_{v'\rightarrow v''}$, and $B_{v'\rightarrow v''}(i)$ for $i=1,2,\dots,7$ are given in tables \ref{fit_VT0}-\ref{fit_VT2}.

\section{Conclusion}

This paper presents a theoretical study of the dissociative recombination 
of BF$^{+}$ and of its competitive processes - vibrational excitation and de-excitation - over a broad range of electron energies - up to about $5$ eV - and considering all the relevant dissociative states within different symmetries. Our MQDT dynamical approach is relying on molecular data calculated by the \textit{ab initio} R-matrix method, completed by 
quantum chemical results.  
The computed Maxwell rate coefficients are relevant for the kinetic modelling of molecule based cold non-equilibrium plasmas, in the context of  complete lack of other theoretical or experimental data on these processes for this
cation.

The present calculations complete our very recent studies of the DR performed on N$^+_2$~\cite{jt591} and on CO$^+$~\cite{mezei2015} 
based on fully {\it ab initio} potential energy curves and couplings 
computed with the R-matrix method \cite{tennyson10}, since BF$^+$ is isoelectronic with these two molecular systems. 
We note that while there is
no experimental data available for the processes considered here for BF$^+$,
our previous calculation on  N$^+_2$ and  CO$^+$ gave good agreement with
the available measurements.

\section*{Acknowledgments}

The authors acknowledge support from the International Atomic Energy Agency (IAEA, Vienna) via the Coordinated Research Project "Light Element Atom, Molecule and Radical Behaviour in the Divertor and Edge Plasma Regions", from the French Agence Nationale de la Recherche via the projects "SUMOSTAI" (No. ANR-09-BLAN-020901) and "HYDRIDES" (No. ANR-12-BS05-0011-01), from the IFRAF-Triangle de la Physique via the project "SpecoRyd", and from the Centre National de la Recherche Scientifique via the programs "Physique et Chimie du Milieu Interstellaire", the PEPS project TPCECAM and the GdR TheM. They also acknowledge generous financial support from R\'egion Haute-Normandie via the CPER, GRR Electronique, Energie et Mat\'eriaux and BIOENGINE project, from the "F\'ed\'erations de Recherche "Energie, Propulsion, Environnement" and "Fusion par Confinement Magn\'etique" ITER, and from the LabEx EMC$^3$, via the project PicoLIBS (No. ANR-12-BS05-0011-01). KC thanks the Institut des Sciences de l'Ing\'enierie et des Syst\`emes (INSIS) of CNRS for a research grant in 2013, and the Laboratoire Ondes et Mat\'eriaux Complexes (LOMC) of Le Havre University for hospitality. SI, NP and IFS acknowledge EU for financial support via the ERASMUS convention between Le Havre University and  Politehnica University of Timisoara. FC and IFS acknowledge EU for financial support via the ERASMUS convention between Le Havre University and University College London and the COST action "Our Astrochemical History".

\section*{Data availability}
Upon a reasonable request, the data supporting this article will be provided by the corresponding author.

%\clearpage

\begin{table*}
\caption{\label{fit_DR}Parameters used in equation (\ref{eqfit_DR}) to represent the DR rate coefficient of BF$^+$. Powers of $10$ are given in square brackets.
}
	  \centering\footnotesize
    \begin{tabular}{lccccccccc}
\hline\hline
    $v$ &  $A_v$  & $\alpha_v$   &  $B_v(1)$ & $B_v(2)$  & $B_v(3)$ & $B_v(4)$    & $B_v(5)$  & $B_v(6)$  & $B_v(7)$ \\
   & & & $\times10^{-4}$ &  $\times10^{-7}$ &  $\times10^{-9}$ &  $\times10^{-11}$ &  $\times10^{-14}$ &  $\times10^{-15}$ &  $\times10^{-17}$ \\
 \hline 
0  &  0.703050632$[-5]$  &  -1.08298227 &  -0.0036398439  &  0.068522753  &  -0.341534791  &  0.86262362  &  -0.117919938  &  0.821543800  &   -0.229194862 \\
1  &  0.205402433$[-11]$  &  0.557133316 &  -0.185580405  &  0.110399946  &  -0.316387400  &  0.531242790  &  -0.0525109108  &  0.283203635  &   -0.064384316 \\
2  &  0.144845145$[-5]$  &  -0.806777388 &  0.285629778  &  -0.322036339  &  1.60926033  &  -4.09338097  & 0.558087964  &  -3.88815002  & 1.08682270 \\
\hline\hline
\end{tabular}
\end{table*}

\begin{table*}
\caption{\label{fit_VT0}Parameters used in equation (\ref{eqfit_VT}) to represent the VE rate coefficient of BF$^+$ 
($v=0$). Powers of $10$ are given in square brackets.
}
	  \centering\footnotesize
    \begin{tabular}{lcccccccccc}
\hline\hline
    $v'\to v''$ & $T_{min}$   &  $A_{v'\to v''}$ & $\alpha_{v'\to v''}$  & $B_{v'\to v''}(1)$   & $B_{v'\to v''}(2)$ & $B_{v'\to v''}(3)$  & $B_{v'\to v''}(4)$  & $B_{v'\to v''}(5)$  & $B_{v'\to v''}(6)$  & $B_{v'\to v''}(7)$ \\
   & (K) & & & $\times10^{-4}$ &  $\times10^{-7}$ &  $\times10^{-10}$ &  $\times10^{-14}$ &  $\times10^{-17}$ &  $\times10^{-19}$ &  $\times10^{-21}$ \\
 \hline 
$0\rightarrow 1$  & 100  & 0.262533924$[-5]$   &  -0.54799987  &   0.2140964    & 0.06393186 &  -0.03384352  &   0.000871079  & -0.00011989 & 0.000084449   &  -0.000023879 \\
$0\rightarrow 2$  & 250  &  0.173038633$[-11]$  &   0.89930301  &   -0.15450686    & 1.03113831 &   -0.80280989 &  0.034503754  & -0.008311545 & 0.01049953  &   -0.00541063 \\
$0\rightarrow 3$  & 350  & 0.228322632$[-19]$  &   2.98753871  &   0.36260816    & -1.87052923 & 4.62415666  &   -0.43742877  &  0.208086587  & -0.4960046  &   0.471734143 \\
$0\rightarrow 4$  & 500  & 0.117098647$[13]$  &  -4.8420988  & 4.40674967    & -5.3514047  &   -0.06486997  &  0.66934409  &  -0.628654524  & 2.37907644 &  -3.33577789 \\
$0\rightarrow 5$  & 700  & 0.409903965$[-3]$  &  -1.1525244  &   1.95259810    & -0.55626919  &   0.845592131 &   -0.25989565  &  0.30064778  & -1.47809013 &   2.6913025 \\
$0\rightarrow 6$  & 1000  & 0.208106654$[-4]$  &  -0.88659016  &  1.64181737   & 1.30228539 &   -4.48925166  &  0.65259363  &   -0.52413341  & 2.27101018 &   -4.13751683 \\
$0\rightarrow 7$  & 1000  & 0.191744424$[-4]$  &  -0.91688598  &  1.76354084  & 0.71583907 &   -2.42161884  &   0.33896208  &   -0.25163098  & 0.96444887 &   -1.47532452 \\
\hline\hline
\end{tabular}
\end{table*}

\begin{table*}
\caption{\label{fit_VT1}Parameters used in equation (\ref{eqfit_VT}) to represent the VE rate coefficient of BF$^+$ 
($v=1$). Powers of $10$ are given in square brackets.
}
	  \centering\footnotesize
    \begin{tabular}{lcccccccccc}
\hline\hline
$v'\to v''$& $T_{min}$ &  $A_{v'\to v''}$ & $\alpha_{v'\to v''}$ & $B_{v'\to v''}(1)$  & $B_{v'\to v''}(2)$  & $B_{v'\to v''}(3)$ & $B_{v'\to v''}(4)$ & $B_{v'\to v''}(5)$ & $B_{v'\to v''}(6)$  & $B_{v'\to v''}(7)$ \\
   & (K) & & & $\times10^{-4}$ &  $\times10^{-7}$ &  $\times10^{-10}$ &  $\times10^{-14}$ &  $\times10^{-17}$ &  $\times10^{-19}$ &  $\times10^{-21}$ \\
 \hline 
$1\rightarrow 0$  & 100  & 0.452471923$[-4]$   &  -0.87073953  &   0.048235539  & -0.015974507 &  0.0020619029 &   -0.000003576 &   -0.00000237 & 0.00000279  &  -0.000001002 \\
$1\rightarrow 2$  & 100  &  0.206894707$[-4]$  &  -0.53410206  &   0.24013867  & 0.024002652 &   -0.011875078 &  0.000294642 &   -0.00003987  & 0.00002774 &   -0.000007757 \\
$1\rightarrow 3$  & 500  & 0.132913428$[-3]$  &   -0.79631192  &   0.52848919  & -0.022729809 & 0.0099201806 &   -0.000246177  &   0.0000335  & -0.00002347 &   0.000006614 \\
$1\rightarrow 4$  & 1000  & 0.618690809$[-3]$  & -1.0599741  &  1.02015058   & -1.02556918 &   2.27295645  &  -0.322366448 &   0.285015517  & -1.42527673 &   3.06796130 \\
$1\rightarrow 5$  & 1000  & 0.195393605$[-5]$  & -0.5544677 &   0.87489847  & 0.410394065 &   -0.950355092 &   0.129851249 &   -0.110278414 & 0.54646568 &   -1.20534084 \\
$1\rightarrow 6$  & 1500  & 0.144480570$[27]$  &  -7.7273169 &   12.1808074  & -69.6758177 &   243.696228 &   -50.8560332 &   63.3554530 & -436.110956 &   1278.90210 \\
\hline\hline
\end{tabular}
\end{table*}

\begin{table*}
\caption{\label{fit_VT2}Parameters used in equation (\ref{eqfit_VT}) to represent the VE rate coefficient of BF$^+$ 
($v=2$). Powers of $10$ are given in square brackets.
}
	  \centering\footnotesize
    \begin{tabular}{lcccccccccc}
\hline\hline
$v'\to v''$& $T_{min}$  &  $A_{v'\to v''}$ & $\alpha_{v'\to v''}$  & $B_{v'\to v''}(1)$ & $B_{v'\to v''}(2)$ & $B_{v'\to v''}(3)$ & $B_{v'\to v''}(4)$ & $B_{v'\to v''}(5)$ & $B_{v'\to v''}(6)$ & $B_{v'\to v''}(7)$ \\
   & (K) & & & $\times10^{-4}$ &  $\times10^{-7}$ &  $\times10^{-10}$ &  $\times10^{-13}$ &  $\times10^{-16}$ &  $\times10^{-18}$ &  $\times10^{-20}$ \\
 \hline 
$2\rightarrow 0$  & 100  & 0.206905711$[-4]$   & -0.53409419 &   -0.005222926  & 0.023974723 &  -0.01186244 &   0.002943012 &   0.000235045 & 0.000277059 &  -0.00007747 \\
$2\rightarrow 1$  & 100  &  0.281749676$[-3]$  & -0.93966107 &   0.05739635  & -0.026671321 &   0.009251884 &  -0.001950454 &   -0.08311545  & -0.000149952 &   0.000039298 \\
$2\rightarrow 3$  & 200  & 0.677540315$[-6]$  &  -0.18052693 &   0.16571344  & 0.078644169 & -0.03551111 &   0.008651454 &   -0.001104037 & 0.000586781 &   -0.000015915 \\
$2\rightarrow 4$  & 500  & 0.312513647$[-2]$  &  -1.13354017 &  0.79154307  & -0.69221004 &   0.88345777 &  -0.684463255 &   0.319179702 & -0.822026586 &   0.896873812 \\
$2\rightarrow 5$  & 1000  & 0.641993202$[-3]$  & -1.01279417 &   0.93152113  & -0.33895961 &   0.140747992 &   0.325809186 &   -0.55855812 & 3.50923558 &   -8.20376383 \\
\hline\hline
\end{tabular}
\end{table*}

\end{document}